\documentstyle[epsfig,twocolumn,aps,prl,epsf,floats]{revtex} %%% Walter

\newcommand{\be}{\begin{equation}}
\newcommand{\ee}{\end{equation}}
\newcommand{\bea}{\begin{eqnarray}}
\newcommand{\eea}{\end{eqnarray}}
\newcommand{\lb}{\left[}
\newcommand{\rb}{\right]}
\newcommand{\lp}{\left(}
\newcommand{\rp}{\right)}

\begin{document}
 
\twocolumn[
\hsize\textwidth\columnwidth\hsize\csname @twocolumnfalse\endcsname
\title{Counting statistics of charge pumping in an open system}
\author{L.S. Levitov}
\address{Physics Department, Center for Materials Science \& Engineering,\\ 
Massachusetts Institute of Technology, Cambridge, MA 02139}
%\date{\today}
 
\widetext %%% Walter

\maketitle

\tightenlines %%% Walter
\widetext  %%% Walter
\advance\leftskip by 57pt  %%% Walter
\advance\rightskip by 57pt  %%% Walter

\begin{abstract}   
  Electron counting statistics of a current pump in an open  
system has universal form in the weak pumping current regime.
In the time domain, charge transmission is described by two 
uncorrelated Poisson processes, corresponding to electron 
transmission in the right and left direction. 
Overall noise is super-poissonian, and can be reduced to poissonian 
by tuning the amplitude and phase of driving signal
so that current to noise ratio is maximized.
Measuring noise in this regime 
provides a new method for determining charge quantum in an open 
system without any fitting parameters.
\vskip2mm
\end{abstract}
]
\bigskip
 
%\begin{multicols}{2}
\narrowtext
 
%%%%%%%%%%%%%%%%%%%%%%%%%%%%%%%%%%%%%%%%%%%%%%%%%%%%%%%%%%%%%%%%%%%%   

% \section{Introduction}
% \label{sec:intro}

Electric current through an open electron system, such as a 
quantum dot well coupled to the leads, can be induced by modulating its area,
shape, or other
parameters\cite{Altshuler'99}.
% \cite{Spivak95,Brouwer98,Zhou1'99,Switkes99,Shutenko00,Aleiner00,Brouwer00,Kravtsov00,Vavilov00,Polianski01}. 
The possibility to generate a DC current through a quantum dot by
cycling potentials on the gates was 
proposed by Spivak et al.\cite{Spivak95} and realized experimentally
by Switkes et al.\cite{Switkes99}. Theory of pumping in open systems
was developed by Brouwer\cite{Brouwer98} and by Zhou et al.\cite{Zhou1'99}.
Brouwer made an interesting observation
that time averaged
pumped current is a purely geometric property of the path in the
scattering matrix parameter space, insensitive to path parameterization 
(also, see Refs.\cite{Avron00,Buttiker'94}).
Zhou et al. demonstrated\cite{Zhou1'99}
that pumping provides a new approach to a detailed understanding of 
mesoscopic transport. Recently, a number of issues related to
incomplete agreement between theory and experiment were 
addressed\cite{Shutenko00,Aleiner00,Brouwer00,Kravtsov00,Vavilov00,Polianski01}. 
An interesting extension of these ideas to mesoscopic
superconducting systems was proposed\cite{Zhou2'99}. 

In this article we discuss
current fluctuations in parametrically driven open systems. 
In the regime of interest, called 
``adiabatic pumping,'' system parameters change slowly
compared to transport time through the system.
This problem is different from adiabatic transport proposed
by Thouless\cite{Thouless'83}, involving an open system with a gap 
in the excitation spectrum. Thouless pump is adiabatic in the 
quantum-mechanical evolution sense, provided that the driving
frequency $f$ is
smaller than the energy gap in the system. 
% So far, this transport mechanism has been realized in single and 
% multiple quantum dots in the Coulomb blockade 
% regime\cite{Kouwenhoven'91,Pothier'92},
% as well as in a point contact driven by an acoustic 
% field\cite{Talyanskii'96}. 
Current in the Thouless pump is quantized in
the units of $ef$, which has been demonstrated in quantum 
dots\cite{Kouwenhoven'91,Pothier'92} and also
motivated proposals to detect fractional Quantum Hall charge\cite{Simon99}. 
% and in Luttinger liquid systems\cite{Simon99,Chamon00}. 
We demonstrate below that, although in
an open system pumped current is not quantized, charge 
quantum can still be detected from noise measurement.

Coherent transport 
through an open mesoscopic system is described\cite{BeenakkerRMP} by
a unitary scattering matrix $S$ which depends on externally driven
parameters, and thus
varies with time. 
The matrix $S(t)$, as a function of time, 
defines a path in the space of all scattering matrices. 
For a system with 
$m$ scattering channels, the matrix space 
is the group $U(m)=SU(m)\times U(1)$. 

In the pumping experiment one can, in principle, realize any path 
in the space of scattering matrices. In this article we consider 
the regime of a weak pumping field, when the path $S(t)$ 
is a small loop. We show that
in this case the distribution of charge $q$ transmitted per cycle
is fully characterized by only two parameters,
average charge flow per cycle, $I=f\langle q\rangle$, 
and noise, $J=f\langle\!\langle q^2\rangle\!\rangle$. 
% The parameters$I$ and $J$ 
% have the meaning of average DC current and noise power, respectively.  
In the time domain charge transport is described 
by {\it two uncorrelated 
Poisson processes} for independent single electron transmission 
to the right and to the left. The generating function for charge 
distribution over 
$N$ pumping cycles in this case is
  \be
  \label{chi-uv}
\chi(\lambda)=\ e^{iuN(e^{i\lambda}-1)} e^{ivN(e^{-i\lambda}-1)},
  \ee
Here the 
rates $u$ and $v$ of right and left transmission per cycle are given by
$u-v=I/ef$, $u+v=J/e^2f$, where 
$e$ is elementary charge. Counting probabilities $p_n$ can be 
found from Fourier decomposition, $\chi(\lambda)=\sum e^{in\lambda} p_n$.  

%%%%%%%%%%%%%%%%%%%%%%%%%%%%%%%%%%%%%%%%%%%%%%%%%%%%%%%%%%%%%%%%%%%%
\begin{figure}
\centerline{\psfig{file=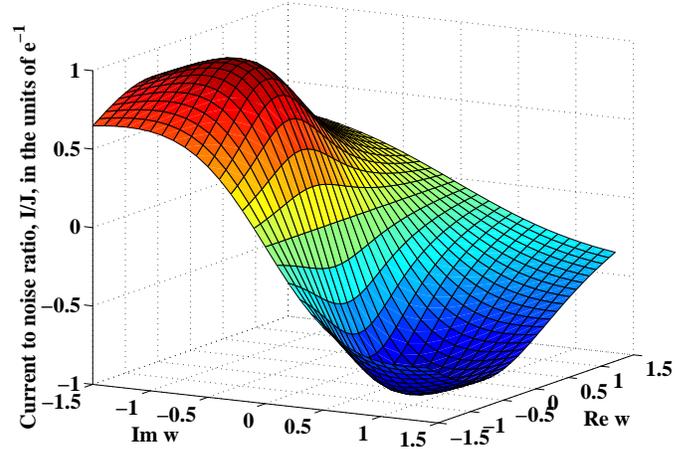,width=3.5in,height=2.5in}}
\vspace{0.5cm}
	\caption[]{
Current to noise ratio, $I/J=e^{-1}(u-v)/(u+v)$,
% $=(\omega/2\pi e)(u-v)/(u+v)$, 
as a function 
of the driving signal parameters (\ref{V12})
for a single channel pump.
The right and left transmission rates $u$ and $v$ 
are given by
(\ref{uv-2channel}). Two harmonic signals driving the pump
are characterized by their 
relative amplitude and phase,
$w=(V_1/V_2)e^{i\theta}$. Maximum and minimum, as a function
of $w$, are $I/J=\pm e^{-1}$, where $e$ is elementary charge.
% Extremum is reached at $-w=z_2/z_1,\ \bar z_2/\bar z_1$.
% The system parameters ratio $z_2/z_1$ used in the plot is $0.6+0.8 i$.
	}    
\label{fig:noiseplot}
\end{figure}
%%%%%%%%%%%%%%%%%%%%%%%%%%%%%%%%%%%%%%%%%%%%%%%%%%%%%%%%%%%%%%%%%%%%

Poisson statistics,
identical to that of conventional classical shot noise, 
makes it possible to use pumped current noise as a new method
of measuring charge quantum. However, 
since right and left current fluctuations (\ref{chi-uv}) are independent, 
one needs a way to separate the two Poisson processes. This can be
achieved, as discussed below, by varying amplitudes and relative phases of 
external driving signals. Either the right or the left transmission
rate, $u$ or $v$, can be nulled. 
The parameters for which this happens give extremum (maximum or minimum)
to the current--to--noise ratio $I/J$ --- see Fig.\ref{fig:noiseplot}. 
Once (\ref{chi-uv}) is reduced to a single
Poisson process, 
this ratio 
gives charge quantum without any fitting parameters,
$J/I=e$. 

So far, only noise in nearly open systems has been 
used to detect quasiparticle charge. In particular,
shot noise measurements in Quantum Hall point 
contacts\cite{de-Picciotto'97,Saminadayar'97}
use backscattering current of 
a conductance plateau. Theoretical discussion of ways
to detect fractional charge in Quantum Hall systems\cite{Kane'94}
and in Luttinger liquid\cite{Bena'00,Chamon00} also
focuses on weak backscattering regime.
Based on the present analysis, we conjecture that
the requirement of ballistic transport is not necessary
if current is induced by weak pumping, rather than by
a DC voltage. The method discussed below  
allows to determine charge quantum 
in open systems with significant scattering.

Another result we report concerns general dependence 
of counting statistics on the path in matrix space. 
  % on ``pumping strategy,'' i.e. 
Different paths $S(t)$, in principle, give rise to different 
current and noise. However, there is a remarkable property 
of invariance with respect to group shifts. Any two paths, 
  \be
  \label{invariance}
S(t)\quad {\rm and}\quad S'(t)=S(t)S_0, 
  \ee
where $S_0$ is a time-independent matrix in $U(m)$, give rise to 
the same counting statistics at zero temperature. 
We note that only the right shifts of the form (\ref{invariance}) leave 
counting statistics invariant, whereas the left shifts generally change it. 
One can explain the result (\ref{invariance}) qualitatively as follows.
The change of scattering matrix, $S(t)\to S'(t)=S(t)S_0$, is equivalent to
replacing states in the {\it incoming} scattering channels by their 
superpositions $\psi^\alpha=S_{0\beta}^\alpha \psi^\beta$. 
At zero temperature, however, Fermi reservoirs are 
noiseless and also such are any their superpositions. Correlation between
superposition states of noiseless reservoirs is negligible, because
all current fluctuations 
arise only during time-dependent scattering. Therefore, noise statistics
remain unchanged. A simple formal proof of the result (\ref{invariance})
is given below. 

 \section{General approach}
 \label{sec:general}

% \nin {\bf General approach:\ }
  % Our discussion will be based on earlier
  % work on noise and statistics\cite{1'93,2'93,3'96,4'96} 
  % in an AC or DC driven system. 
The distribution of transmitted charge 
is characterized\cite{1'93,2'93} by electron counting probabilities 
$p_n$, usually accumulated in one generating function 
$\chi(\lambda)=\sum e^{in\lambda}p_n$. The function
$\chi(\lambda)$ is given by Keldysh partition function,
describing evolution in the presence of a counting field $\lambda$ which is 
an auxiliary gauge field having opposite signs on the forward and backward
parts of Keldysh time contour\cite{3'96,4'96}. 
By a gauge transformation, 
the time-dependent scattering matrix becomes  
$S_{\lambda}(t)=e^{i\frac{\lambda}4\sigma_3}S(t)e^{-i\frac{\lambda}4\sigma_3}$
for the forward time direction, and $S^\dagger_{-\lambda}(t)=
e^{-i\frac{\lambda}4\sigma_3}S^\dagger(t)e^{i\frac{\lambda}4\sigma_3}$ for the 
backward direction, where $\sigma_3$ is a diagonal matrix
with eigenvalues $1$ and $-1$ for the right and left channels. 
Then 
% the generating function $\chi(\lambda)$ is given 
% by the following formula:
  \bea
\label{chi-det}
  \chi(\lambda)=\ {\rm det}\lp \openone +
n(t,t') \lp {\hat T}_\lambda(t)-\openone\rp \rp\\
\label{T-S}
{\hat T}_\lambda(t)=S^\dagger_{-\lambda}(t)S_{\lambda}(t),\qquad
n(t,t')=\frac{i}{2\pi(t-t'+i\delta)}
% S_{\lambda}(t)=e^{-i\frac{\lambda}4\sigma_3}S(t)e^{i\frac{\lambda}4\sigma_3}
  \eea
where $\hat n$ is the density matrix of reservoirs 
at zero temperature. The determinant of an infinite matrix
(\ref{chi-det}) requires proper understanding and 
definition\cite{3'96,4'96}.

There are two ways of handling the determinant (\ref{chi-det}). 
For periodic $S(t)$, one can use frequency representation\cite{2'93}
in which $\hat n$ is a diagonal
operator, $n(\omega)=\theta(-\omega)$. The 
operator $S(t)$ has off-diagonal
matrix elements 
$S_{\omega',\omega}$ with frequency change being a multiple
of external (pumping) frequency, $\omega'-\omega=n\Omega=2\pi nf$. In this method the energy
axis is divided into intervals $n\Omega<\omega<(n+1)\Omega$, and each interval 
is treated as a separate 
conduction channel. In doing so it is convenient (and some times necessary) 
to assign separate
counting field $\lambda$ to each frequency channel, so that the field
$\lambda$ may acquire frequency 
channel index in addition to the usual conduction channel dependence given by $\sigma_3$ in
(\ref{chi-det}). This procedure brings (\ref{chi-det}) to the form of a determinant 
of a matrix with an infinite number of rows and columns. This matrix is then
truncated at very high and low frequencies, eliminating 
empty states and the states deep in the Fermi sea which do not contribute to 
noise. 

This method was used in Ref.\cite{2'93} to study noise 
in a two channel problem described by 
a $2\times2$ matrix 
  \be
  \label{matrix-93}
S(\tau)\equiv\lp\matrix{r & t' \cr t & r' }\rp
=\lp\matrix{B+be^{-i\Omega \tau} & \bar A+\bar ae^{i\Omega \tau} \cr 
A+ae^{-i\Omega \tau} & -\bar B-\bar be^{i\Omega \tau} }\rp
  \ee
which is unitary for $|A|^2+|a|^2+|B|^2+|b|^2=1$, $A\bar a+B\bar b=0$. 
Charge distribution for $N$ pumping cycles is 
  \be
  \label{chi-matrix-93}
\chi(\lambda)=\lp 1+p_1 (e^{i\lambda}-1)+p_2 (e^{-i\lambda}-1) \rp^N 
  \ee
with $p_1=|a|^4/(|a|^2+|b|^2)$ and $p_2=|b|^4/(|a|^2+|b|^2)$. 

Alternatively, the determinant in (\ref{chi-det}) can be analyzed in the time domain. 
% Although this method is less general than that of Ref.\cite{2'93}, 
% it is quite useful for certain forms of the time dependence $S(t)$. 
This representation is beneficial
when an orthogonal basis of functions can be found that provides
a simple enough representation of  
(\ref{chi-det}). For dealing with periodically driven systems, it is convenient
to close the time contour by imposing antiperiodic boundary conditions on the 
interval $0<t<\tau_0\equiv 2\pi N/\Omega$, where $N$ is the number of pumping cycles. 
Upon doing this, the reservoir density matrix $\hat n$ 
acquires the form
  \be
  \label{n-periodic}
n(t,t')=\frac{i}{2 \tau_0\ \sin\lp \pi(t-t'+i\delta)/\tau_0\rp}
  \ee
The reason for introducing periodicity (\ref{n-periodic}) is that 
in this problem one deals with probability distributions which are
stationary in time, so that $\chi(\lambda)$ is multiplicative, 
and $\ln\chi(\lambda)\propto N$ at $N\gg 1$. Conveniently, 
the periodicity of (\ref{n-periodic}) makes multiplicative character
of $\chi(\lambda)$ an exact property, true even for $N\simeq 1$.
% We use periodic time representation below.

One encounters a number of interesting cases which can be handled 
in the time domain in the problem of noise in voltage-driven systems.
An external driving voltage $V(t)$ applied across 
the system can be incorporated in $S$ as a time-dependent forward scattering phase. 
This is achieved by a gauge transformation making the scattering 
matrix time-dependent:
  \be
  \label{V-path}
S(t)=e^{-\frac{i}2\varphi(t)\sigma_3}Se^{\frac{i}2\varphi(t)\sigma_3},
\quad 
\dot\varphi(t)=\frac{e}{\hbar}V(t)
\ .
  \ee
The formula (\ref{V-path}) defines a circular path in $U(m)$
of radius which depends on the system conductance. 
Full statistics have been studied for a large family of paths 
of the form (\ref{V-path}). The statistics was found to be binomial 
for the DC voltage case\cite{1'93} as well as for the AC case\cite{3'96,4'96} 
with a particular time dependence $V(t)$ obtained from the criterion 
of minimal noise.

An attempt to adapt these results to the problem of pumping noise 
was made by Andreev and Kamenev\cite{AndreevKamenev}. For several 
matrix paths $S(t)\in SU(2)$ obtained from (\ref{V-path}) by exchanging 
incoming scattering channels, 
binomial charge distributions arise, not surprisingly, with transmission 
and reflection probabilities exchanged. Another matrix considered in
Ref.\cite{AndreevKamenev}, 
$r=-r'=\cos\Omega t$, $t=t'=\sin\Omega t$, is related to (\ref{matrix-93}) 
by the transformation
(\ref{invariance}) with $S_0=(\sigma_3+\sigma_2)/\sqrt{2}$ and 
parameter values $a,b=1/\sqrt{2}$, $A,B=0$. The result 
(Eq.8 of Ref.\cite{AndreevKamenev}) in this case agrees
with (\ref{chi-matrix-93}), in full accord with the invariance property
(\ref{invariance}). 
However, in the context of the pumping problem, 
the paths $S(t)$ considered 
in Ref.\cite{AndreevKamenev} appear to be less relevant than, say, the paths
(\ref{V-path}) in voltage-driven systems. In general,
the dependence of 
the scattering matrix $S(t)$ on the parameters externally controlled
in the pumping experiment is not known. 
Because of that, the results for particular paths $S$ are of less interest
than the properties that hold for sufficiently general families
of paths. 

 \section{Calculation}
 \label{sec:method}

% \nin {\bf Calculation:\ }
For a weak pumping field we shall evaluate (\ref{chi-det}) 
in the time domain 
by expanding $\ln {\rm det}(...)$ in powers of $\delta S$ and keeping
non-vanishing terms of lowest order. In doing so, however, we preserve
full functional dependence on $\lambda$ which gives
all moments of counting statistics. We write
$S(t)=e^{A(t)}S^{(0)}$ with antihermitian $A(t)$ 
representing small perturbation, ${\rm tr} A^\dagger A\ll1$.
Here $S^{(0)}$ is scattering matrix of the system in the absence of 
pumping. Substituting this into (\ref{T-S}) one obtains
  \be
\label{T-lambda}
{\hat T}_\lambda(t)\equiv
{\hat T}_\lambda^{(0)}+\delta T_\lambda(t)=
S^{(0)\dagger}_{-\lambda} e^{-A_{-\lambda}(t)}e^{A_{\lambda}(t)}S^{(0)}_{\lambda}
  \ee
with ${\hat T}_\lambda^{(0)}=S^{(0)\dagger}_{-\lambda} S^{(0)}_{\lambda}$
and
$A_{\lambda}(t)=
e^{i\frac{\lambda}4\sigma_3}A(t)e^{-i\frac{\lambda}4\sigma_3}$. 
Now, we expand (\ref{chi-det}):
  \be
\label{ln-chi}
\ln \chi(\lambda)=\ln {\rm det} Q_{0}+ 
{\rm tr} R
-\frac12 {\rm tr} R^2
+\frac13 {\rm tr} R^3
-...
  \ee
where $Q_{0}=1+\hat n({\hat T}_\lambda^{(0)}-1)$
and $R=Q_{0}^{-1}\hat n\delta T_{\lambda}$. 
At zero temperature, from $\hat n^2=\hat n$ it follows that
${\rm det} Q_{0}=1$ 
and $R=S^{(0)\!\ -1}_{\lambda}
\hat n \lp e^{-A_{-\lambda}(t)}e^{A_{\lambda}(t)}-1\rp S^{(0)}_{\lambda}$. 
Therefore,
% Hence (\ref{ln-chi}) becomes
  \be
\label{ln-M}
\ln \chi(\lambda)= {\rm tr}\ \hat n\hat M
-\frac12 {\rm tr} (\hat n\hat M)^2
+\frac13 {\rm tr} (\hat n\hat M)^3
-...
  \ee
where $\hat M=e^{-A_{-\lambda}(t)}e^{A_{\lambda}(t)}-1$. 
Note that at this stage there is no dependence left on
the constant matrix $S^{(0)}$, which proves invariance 
under the group shifts (\ref{invariance}). 

We need to expand (\ref{ln-M})
in powers of the pumping field, which amounts to taking 
the lowest order terms of the expansion in powers of
the matrix $A(t)$. One can check that the two ${\cal O}(A)$ terms 
arising 
from the first term on the RHS of (\ref{ln-M}) vanish. 
The ${\cal O}(A^2)$ terms 
arise from the first and 
second term in (\ref{ln-M}) and have the form
   \be
\label{AB}
\ln \chi= %(\lambda)= 
\frac12 {\rm tr}\lp \hat n \lp
A_{-\lambda}^2 +A_{\lambda}^2 -\! 2 A_{-\lambda}A_{\lambda}\rp\rp
\! -\frac12 {\rm tr} (\hat n B_{\lambda})^2
  \ee
with $B_{\lambda}(t)=A_{\lambda}(t)-A_{-\lambda}(t)$. 
At zero temperature, by using 
$\hat n^2=\hat n$, one can bring (\ref{AB}) to the form
   \be
\label{ln-AB}
%\ln \chi(\lambda)= 
\frac12\!\ {\rm tr}\!\ \lp\hat n 
\lb A_{\lambda},A_{-\lambda}\rb\rp
+\frac12 \lp {\rm tr} \lp \hat n^2 B_{\lambda}^2 \rp
- {\rm tr} (\hat n B_{\lambda})^2 \rp
  \ee
The first term of (\ref{ln-AB}) has to be regularized in 
the Schwinger anomaly fashion, by splitting points, 
$t',t''=t\pm \epsilon/2$, which gives
   \be
   \label{anomaly}
\frac12 \oint n(t',t'')
{\rm tr} \lp A_{-\lambda}(t'') A_{\lambda}(t')
-A_{\lambda}(t'') A_{-\lambda}(t') \rp
dt 
% = \int {\rm tr}\lb \dot A_{\lambda}(t), A_{-\lambda}(t)\rb dt,
  \ee
% with $t',t''=t\pm \epsilon/2$. 
Averaging over small $\epsilon$ can be achieved either by
inserting in (\ref{anomaly}) additional integrals over 
$t'$, $t''$, or simply by replacing 
$A_{\lambda}(t)\to \frac12\lp A_{\lambda}(t)+A_{\lambda}(t')\rp$, etc.
After taking the limit $\epsilon\to 0$,
Eq.(\ref{anomaly}) becomes
   \be
\label{int1}
\frac{i}{8\pi} \oint {\rm tr}\lp 
A_{-\lambda}\partial_t A_{\lambda}
-
A_{\lambda}\partial_t A_{-\lambda}
\rp dt
% -\frac12 \int\!\!\int 
% \frac{{\rm tr} \lp B_{\lambda}(t)-B_{\lambda}(t')\rp^2}{(t-t')^2} dt dt'
   \ee
The second term of (\ref{ln-AB}) can be written as
   \be
\label{int2}
\frac1{4(2\pi)^2} \oint\!\!\oint 
\frac{{\rm tr} \lp B_{\lambda}(t)-B_{\lambda}(t')\rp^2}{(t-t')^2} dt dt'
   \ee
Now, we decompose $A=a_0+z+z^\dagger$,
so that $\lb \sigma_3, a_0\rb =0$, 
$\lb \sigma_3, z\rb =-2 z$,
$\lb \sigma_3, z^\dagger\rb =2 z^\dagger$. Then 
  \bea
A_\lambda\equiv
e^{-i\frac{\lambda}4\sigma_3} A e^{i\frac{\lambda}4\sigma_3}
=a_0+e^{i\frac{\lambda}2}z^\dagger+e^{-i\frac{\lambda}2}z
\\
B_\lambda=\lp e^{i\frac{\lambda}2}-e^{-i\frac{\lambda}2} \rp
W,\quad W\equiv z^\dagger-z
  \eea
Substituting this into (\ref{int1}) and (\ref{int2}) one finds that 
in terms of $W(t)$ these two expressions become
  \be
\label{term1}
\frac{\sin\lambda}{8\pi}\oint {\rm tr} \lp \lb \sigma_3,W\rb \partial_t W\rp dt
  \ee
and
  \be
\label{term2}
\frac{\lp 1-\cos\lambda\rp}{2(2\pi)^2}\oint\!\!\oint 
\frac{{\rm tr} \lp W(t)-W(t')\rp^2}{(t-t')^2} dt dt'
  \ee
Hence $\ln\chi$ indeed depends on $\lambda$ as
$u(e^{i\lambda}-1)+v(e^{-i\lambda}-1)$. 

Eq.(\ref{term1}) is essentially identical to the result obtained by 
Brouwer for average pumped current\cite{Brouwer98}. The integral
in (\ref{term1}) is invariant under reparameterization, and thus 
has a purely geometric character determined by the contour $S(t)$
in $U(m)$. Eq.(\ref{term2}) represents a generalization of the 
expression for noise induced by time-dependent external field
considered in Refs.\cite{unpublished'93,3'96}.

The parameters $u$ and $v$ in (\ref{chi-uv}) can be expressed 
through $z(t)$ and $z^\dagger(t)$ in a simple way. 
Let us write $z(t)$ as $z_{+}(t)+z_{-}(t)$,
where $z_{+}(t)$ and $z_{-}(t)$ contain only 
positive or negative Fourier harmonics, respectively. Then
  \bea
\label{u-v}
u=\frac{i}{4\pi}\oint {\rm tr}\lp 
z^\dagger_{-}\partial_t z_{+} -
z_{+} \partial_t z^\dagger_{-}
\rp dt
=\sum\limits_{\omega>0} \omega\ {\rm tr} z^\dagger_{-\omega} z_{\omega}
,\\
v=\frac{i}{4\pi}\oint {\rm tr}\lp z_{-} \partial_t z^\dagger_{+}
-  z^\dagger_{+}\partial_t z_{-}
\rp dt
=-\sum\limits_{\omega<0} \omega\ {\rm tr} z^\dagger_{-\omega} z_{\omega}
  \eea
Note that $u\ge0$ and $v\ge0$.
It is straightforward to show that (\ref{term1}) equals $i\sin\lambda (u-v)$,
whereas (\ref{term2}) equals $(\cos\lambda-1) (u+v)$, 
which completes the proof of (\ref{chi-uv}).

% \section{Detecting charge quantum in an open system}

Now we consider a single channel pump, $S(t)\in U(2)$. In this case, 
$z$ and $z^\dagger$ are complex numbers.
For harmonic driving signal, without loss of generality, one can write
  \be
  \label{V12}
z(t)=z_1V_1\cos(\Omega t+\theta)+z_2V_2\cos(\Omega t)
,
  \ee
where $V_{1,2}$ are pumping signal amplitudes, and complex parameters 
$z_{1,2}$ depend on microscopic details. From (\ref{u-v}) we find 
the Poisson rates
  \be
  \label{uv-2channel}
u=\frac14 \left| z_1V_1e^{i\theta}+z_2V_2 \right|
,\ 
v=\frac14 \left| z_1V_1e^{-i\theta}+z_2V_2 \right|
  \ee
Note that $u$ and $v$ vanish at particular signal amplitudes 
ratio $V_1/V_2$ and phase $\theta$. Once the two Poisson processes
(\ref{chi-uv}) are reduced to one, the current to noise ratio
gives elementary charge, $I/J=\pm e^{-1}$. This happens at
the extrema of $I/J$ as a function of $w=(V_1/V_2)e^{i\theta}$, 
for (\ref{uv-2channel}) reached
at $w=-z_2/z_1,\ -\bar z_2/\bar z_1$.
This behavior is illustrated in Fig.\ref{fig:noiseplot}, 
where $z_2/z_1=0.4+0.8 i$ is used. 

Reducing the counting statistics (\ref{chi-uv}) to purely Poissonian 
by varying pumping signal parameters, in principle,
is possible for any number of channels $n$.
However, since the number of parameters to be tuned is $2 n^2$, 
this method is practical perhaps in the single channel case only. 
Although the method is demonstrated for non-interacting 
fermions, we argue that it can be applied to
interacting systems as well. Poisson distribution results from the absence 
of correlations of subsequently transmitted particles, which must be the 
case in any system at small pumping current. Using the dependence of  
Poisson rates $u$, $v$ on the driving signal to maximize $I/J$
allows to eliminate one of the two Poisson processes (\ref{chi-uv}) and then 
obtain charge quantum in the standard way as $e=J/I$.

%%%%% acknowledgement

I am grateful to Dmitry Novikov for useful discussion. 
This work was supported by the MRSEC Program of the National Science 
Foundation under Grant No. DMR 98-08941

%\end{multicols}
\end{document}